\documentclass[aps,pre,review,showpacs,superscriptaddress,12pt]{revtex4-2}
\usepackage{amssymb,amsmath}
\usepackage{amsfonts}
\usepackage{dcolumn}
\usepackage{lipsum}
\usepackage{mathrsfs}
\usepackage{graphicx,subfigure}
\usepackage{color}
\usepackage{comment}
\usepackage{epsfig}
\usepackage{rotating}
\usepackage{tikz}

\newcommand{\iisc}
{\affiliation{Centre for Condensed Matter Theory, Department of Physics, Indian Institute of Science, Bangalore 560012, India}}
\newcommand{\ongil}
{\affiliation{ONGIL.AI, Chennai 600113, India}}

\newcommand{\A}
{\affiliation{Interdisciplinary Mathematical Sciences, Department of Mathematics, Indian Institute of Science, Bangalore 560012, India}}

\newcommand{\B}
{\affiliation{Bioinformatics Lab, Department of Biochemistry, Indian Institute of Science, Bangalore 560012, India}}

\newcommand{\ibm}
{\affiliation{IBM Research Labs, Bangalore 560045, India}}

\newcommand{\icts}
{\affiliation{International Center for Theoretical Sciences, TIFR, Bangalore 560089, India}}

\begin{document}
\title{Percolation in a three-dimensional non-symmetric multi-color loop model}
\date{\today}

\author{Soumya Kanti Ganguly}%
\email[Email:]{gangulysoumyakanti@gmail.com}
\iisc
\email[Email:]{skg@ongil.ai}
\ongil

\author{Sumanta Mukherjee}%
\email[Email:]{sumantamukherjee@ibm.com}
\A\B\ibm

\author{Chandan Dasgupta}%
\email[Email:]{cdgupta@iisc.ernet.in}
\iisc\icts

\begin{abstract}
We conducted Monte Carlo simulations to analyze the percolation transition of a non-symmetric loop model 
on a regular three-dimensional lattice. We calculated the critical exponents for the percolation transition 
of this model. The percolation transition occurs at a temperature that is close to, but not exactly the 
thermal critical temperature. Our finite-size study on this model yielded a correlation length exponent 
that agrees with that of the three-dimensional XY model with an error margin of six per cent.
\end{abstract}

\maketitle

\section{Introduction}
It has been observed that porous materials like charcoal, limestone or sponge are impervious to water until a substantial fraction 
of the pores are filled by the same. The threshold value of these occupied pores is a measure of a kind of phase transition popularly 
known as percolation. This threshold value that separates the two phases and is known as the \textsl{percolation threshold}. Perhaps, 
the most influential work in this field which laid the mathematical foundations of this subject was the work by Broadbent and Hammersley 
\cite{Broadbent}, in 1957. Following their work, there has been an avalanche of activities in both physics and the field of applied 
sciences, which continues even today. One of the early applications of the theory was due to P.W. Anderson, where he used percolation 
to study the localization of electrons in disordered media, a phenomenon which is famously known as Anderson Localization \cite{Anderson}. 
Subsequently, people used it to study the conduction and transport phenomena in inhomogeneous conductors near the percolation threshold 
\cite{Thouless,Kirkpatrick1,Shklovski}. In statistical mechanics, Fisher and Essam studied percolation in exotic lattice systems like the 
Bethe lattice (infinite Cayley tree) \cite{Fisher}. But it was not until the late 1960s when Fortuin and Kasteleyn \cite{Kasteleyn,Fortuin} 
for the first time showed that there exists a direct connection between the percolation problem and the Ising or Ising-like models. This 
connection was vital in calculating the critical exponents associated with the percolation transition \cite{Essam,Harris,Young,Kirkpatrick2}. 
The central idea was the concept of \textsl{clustering}, which was later exploited by physicists to circumvent the problem of critical slow 
down in the Monte Carlo simulations of spin systems \cite{Hoshen,Swendsen,Wolff}. The ideas of percolation have also been used in the study 
of non-physical phenomena e.g. forest fires \cite{Stauffer}, interactions in social networks \cite{Newman}, and many others. 

Compared to the conventional site and bond percolation, the phenomenon of loop percolation is a relatively less studied subject in the 
physics community. Especially when the percolation transition is thermally driven. The statistical distribution of these loops can be 
thought of as clusters or connected components, whose growth has a very natural description in terms of the percolation transition 
\cite{Strobl,Williams1,Williams2}. For example, the onset of the $\lambda$-transition in Superfluid Helium is mediated by the percolation 
of vortex rings \cite{Williams1}. Extensive studies of these condensed matter systems have also led to our understanding of the formation 
of topological defects in the universe \cite{Kibble,Zurek}. A statistical mechanical study of the frustrated XY model by Nguyen and Sudbo 
\cite{Sudbo}, has shown that the temperature-driven percolation transition due to the vortices was responsible for the melting of the Abrikosov 
flux-line lattice in type-II superconductors. In the case where temperature was not the driving factor, a study by Pfeiffer and Rieger 
\cite{Pfeiffer} showed that the critical exponents for the loop percolation transition belong to the same universality class as the conventional 
bond or site percolation. However, in the same study, the percolation transition which occurs due to minimizing the ground state ($T = 0$) 
energy of a loop Hamiltonian with random disorder, resulted in a completely different set of critical exponents. It is worth mentioning that 
the work by Pfeiffer and Rieger partly serves as the motivation for the present work. To the best of our knowledge, an extensive study of the 
universal behavior of thermally driven percolating loops remains unaddressed to this date. Therefore, we believe that there is further scope 
for work and our study will be a step in that direction.

In this paper, we look at the finite temperature percolation transition observed in a certain loop model, in a regular three-dimensional 
lattice. The model under consideration is a variant of the loop Hamiltonian used in \cite{Pfeiffer}. The major differences between our loop 
model and the one in \cite{Pfeiffer} is that our loop model Hamiltonian is devoid of any random disorders, and it has multiple loop components 
where each component is termed as a \textsl{color}. We will refer to our model as the non-symmetric or NS loop model, for reasons that have been 
discussed in our companion paper (Ganguly et al \cite{Ganguly}, hereafter GMDa24). This paper contains the thermodynamic properties of the NS loop 
model, including its thermal critical exponent (specific heat $\alpha$), and the correlation length exponent ($\nu$). In the present work, we have 
calculated the critical exponents for percolation transition seen in the NS loop model. We observe that the percolation transition happens 
at a temperature close to but not equal to the thermal critical temperature. Our finite-size scaling analysis suggests that the correlation length 
exponent for the NS loops is that of the XY model in three dimensions. Since the loop percolation transition in NS loops is a thermally driven 
transition, its percolation properties are intertwined with its thermal properties. Therefore, it is important to discuss the finite temperature 
properties of the NS loop model before we discuss its percolation properties. 

The rest of the paper is organized as follows. In Section II, we introduce the NS loop model and briefly discuss its finite temperature properties. 
In Section III, we discuss the percolation transition in the NS loop model using finite-size scaling. Finally, we end with some concluding remarks in 
Section IV.

\section{The Multi-colored loop model}
\begin{figure}
\centering
\subfigure[]{\includegraphics[width=0.55\textwidth]{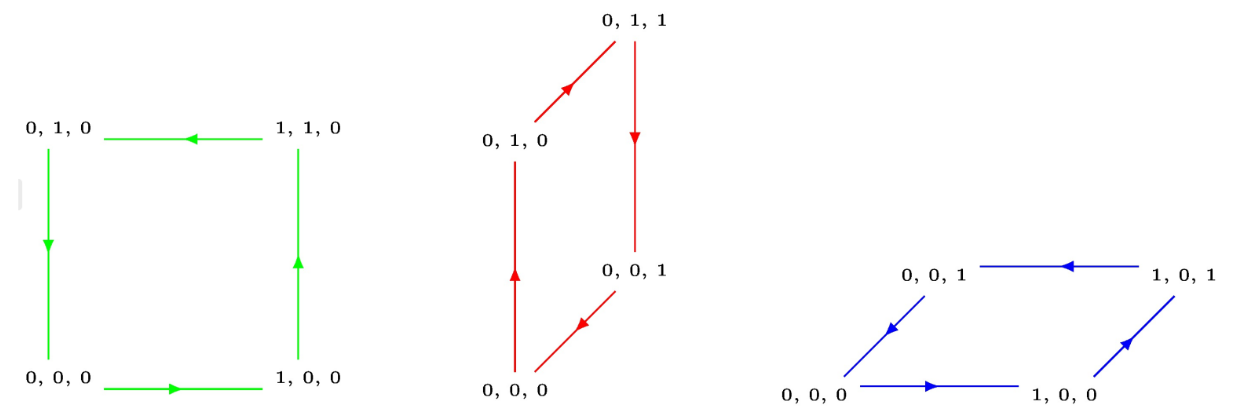}}
\subfigure[]{\includegraphics[width=0.9\textwidth]{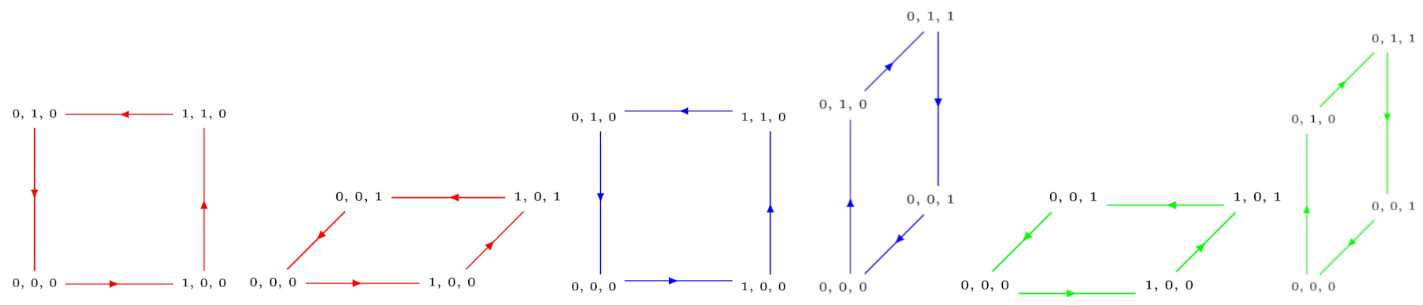}}
\subfigure[]{\includegraphics[width=0.54\textwidth]{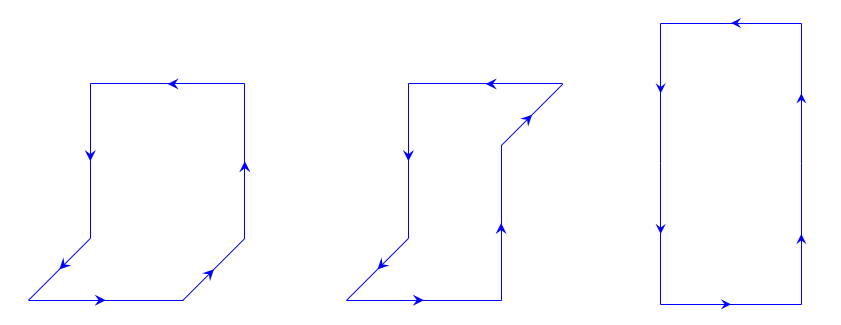}}
\caption{Elementary excitations in NS-loops: (a) shows the first excited state with 1 unit of energy ($k_{B}T$ units) and degeneracy 6;
(b) shows the second excited state with 1.5 units of energy and degeneracy 6; (c) shows the third excited state (left and middle) 
with 2 units of energy, and the fourth excited state (right) with 2.5 units of energy respectively.}
\label{NSLoopExcitations12}
\end{figure}
The order-disorder transitions in a regular three-dimensional lattice system can occur due to topological defects. Their topological 
nature protects them and prevents their removal by any smooth deformation of the lattice. They obey a continuity condition 
(see Eq.\eqref{continuity}), for which their lattice realizations assume the form of closed loops (Fig.\ref{NSLoopExcitations12}). 
In solids, these topological defects are called dislocations and disclinations. The former is responsible for the breakdown of the 
broken translational symmetry, while the latter destroys the broken rotational symmetry \cite{Kleinert1,Kleinert2,Kleinert3}. 
The topological defects in solids are second-rank symmetric tensors, which can cause melting. In three-dimensional solids, the melting transition 
is a first-order transition. In our companion paper (GMDa24), we have argued the theoretical possibility of tensor loop defects which can be 
non-symmetric, hence the name NS loops. The NS loops undergo a second-order phase transition, but we have shown that strong interactions among 
the various colors can alter the nature of the transition in these loop systems. In this section, we will briefly discuss the NS loop model and 
refer to GMDa24 for more details.

\begin{figure}
\centering
{\includegraphics[width=0.5\textwidth]{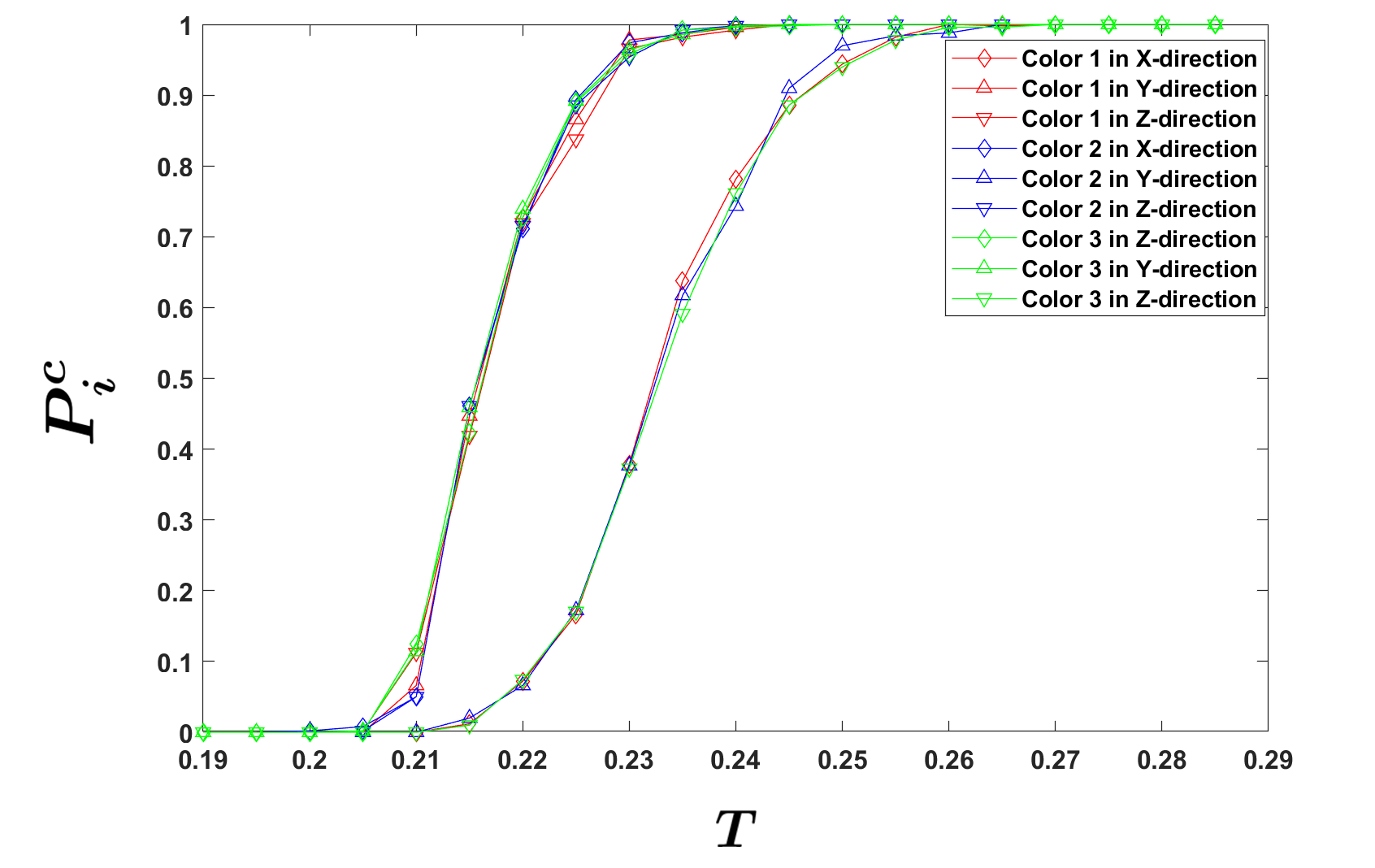}}
\caption{Percolation probability among different colors ($c$) in different directions ($i$) (size $L = 16$).}
\label{PercolationProbabilityColorsNonSymmLoops}
\end{figure}

Let $\eta_{ij}(\mathbf{x})$ represent the integer-valued NS tensors at each point $\mathbf{x}$ in the lattice. The tensor has nine independent 
elements, with each column denoted by index $j$, which corresponds to the color of the loop (e.g. $j \in \{1,2,3\} = \{red,blue,green\}$). For 
each color, we have the following lattice continuity equation given by
\begin{equation}\label{continuity}
\Delta_{i}\eta_{ij}(\mathbf{x}) = 0 \quad j \in \{1,2,3\},
\end{equation}
The Hamiltonian, which is a function of the $\eta_{ij}$'s is given by
\begin{eqnarray}\label{NSLoopHamiltonian}
H(\eta_{ij})_{loop} = \sum_{\mathbf{x}}
  &A&\Big[ \eta^{2}_{11}(\mathbf{x}) + \eta^{2}_{22}(\mathbf{x}) + \eta^{2}_{33}(\mathbf{x}) \Big]  \\ \nonumber
+ &B&\Big[ \eta^{2}_{12}(\mathbf{x}) + \eta^{2}_{21}(\mathbf{x}) + \eta^{2}_{23}(\mathbf{x})        \\ \nonumber
  &+&\eta^{2}_{32}(\mathbf{x}) + \eta^{2}_{13}(\mathbf{x}) + \eta^{2}_{31}(\mathbf{x}) \Big]        \\ \nonumber
+ &D&\Big[ \eta_{11}(\mathbf{x})\eta_{22}(\mathbf{x}) + \eta_{22}(\mathbf{x})\eta_{33}(\mathbf{x}) \\ \nonumber
  &+& \eta_{33}(\mathbf{x})\eta_{11}(\mathbf{x}) \Big],
\end{eqnarray}
where $A = 0.5, B = 0.25$ and $D = 0.1$, in units of $k_{B}T$. If $E^{(\textrm{Color})}_{(\textrm{Plane})}$ denote the energy of an elementary 
excitation of color $c$ along a given plane, then the excitations in Fig.\ref{NSLoopExcitations12}(a),(b) will have energies given by
\begin{eqnarray}\label{Elementary_excitation2}
&&E^{1}_{YZ} = E^{2}_{XZ} = E^{3}_{XY} = 4B, \\ \nonumber
\textrm{and} \quad &&E^{1}_{XY} = E^{1}_{XZ} = E^{2}_{XY} = E^{2}_{YZ} = E^{3}_{YZ} = E^{3}_{XZ} = 2(A+B). \\ \nonumber
\end{eqnarray}
Where, $4B < 2(A+B)$. From the calculations above, it is evident that the excitations carrying $4B$ units of energy have greater chances of 
acceptance than the ones carrying $2(A + B)$ units of energy. As a result, color-1 in the Y, and Z directions, color-2 along X, and Z directions 
and color-3 in the X, and Y directions have their corresponding percolation probabilities shooting up close to the criticality as shown in 
Fig.\ref{PercolationProbabilityColorsNonSymmLoops}. We must emphasize that despite such directional preferences, the critical behavior of this 
model and the model with complete isotropy are the same.
\section{Percolation transition in NS loops}
In the previous section, we have seen that the basic excitations in the NS loop system are elementary loops (plaquettes). 
They can be oriented along any arbitrary plane and can have both clockwise and anti-clockwise \textsl{chiralities}. Every site associated 
with such a loop will have an incoming and an outgoing arrow whose number is always conserved. Near the \textsl{critical point}, when the 
system is proliferated by such loopy networks, the system is said to be disordered and there will exist a path or multiple paths connecting 
the opposite sides of the system running through the bulk. The analogy between site percolation and percolation due to loops stems from the 
following fact. A single loop (Fig.\ref{NSLoopExcitations12}(a)), which is the most basic excitation in the system, is a connected component 
of size four. That is, it connects four sites in the lattice. As the number of loops increases in the system, they will join one another to 
form bigger loops or clusters (Fig.\ref{NSLoopExcitations12}(c)). Near the percolation transition, the typical size of the connected components 
will grow until we end up having a percolating cluster. This implies that we can always find a path or a set of paths within the cluster that 
run through the bulk of the system connecting its opposite faces (boundaries). In other words, we will have loops whose \textsl{diameter} will 
scale as the size of the system when the system is driven near the critical region. Unlike conventional phase transitions, this kind of transition 
is a thermally driven geometrical transition. It is a continuous transition which involves occupation of the sites (site percolation) or bonds 
(bond percolation) in a lattice, and therefore will have its own critical exponents.

With increasing temperature these loops interact with one another to form clusters or connected components of varying sizes. If two points within 
the system are connected by a path, then they belong to the same cluster. If these two points happen to be at the two opposite boundaries or faces 
of the system, then we say that we have a percolating path and the cluster is a percolating cluster. We will see that near the critical point, the 
number of such paths and the size of the percolating cluster have a scaling behavior. Our knowledge of the statistics of the connected components 
(loop statistics) provides information about the geometry of these networks and enables us to calculate the critical exponents. Contrary to ordinary 
percolation, where bonds/sites are randomly and independently occupied (or emptied) with some probability, the thermal problem will have them filled 
or emptied based on the Metropolis algorithm. The algorithm's conditional addition or removal of bonds/sites puts it in a special category of problems 
in statistical mechanics known as \textsl{correlated percolation}. They are correlated because the loops are generated by a first-order Markov chain of 
events.

\subsection{Method}

Our goal is to investigate the percolation transition in NS loops within a regular three-dimensional lattice with \textsl{periodic boundary conditions} 
using computer simulations. As previously stated, the loop variables are integer-valued quantities that obey a continuity condition. Therefore, the update 
equations must satisfy this constraint (GMDa24)\cite{Ganguly}. Since the percolation transition is thermally driven, we employ the Metropolis algorithm 
\cite{Murthy,Binder} to simulate the Hamiltonian in Eq.\eqref{NSLoopHamiltonian} at a finite temperature $T$. According to the algorithm, the evolution of the 
system of loops is a Markov process. If $\Delta E$ represents the change in the energy of the system when transitioning from state $a$ to state $b$, then 
the \textsl{transition probability} $W(a \rightarrow b)$ is given by 

\begin{equation}\label{Metropolis1}
W(a \rightarrow b) = \begin{cases}
  \begin{array}{c} 
	  \exp(-\Delta E/k_{B}T) \\
          1
  \end{array}
  & 
  \begin{array}{c} 
  \textrm{if} \quad \Delta E > 0, \\ 
  \textrm{otherwise.} 
 \end{array}
 \end{cases}
\end{equation}

After the system has equilibrated at temperature $T$, we will have a canonical ensemble of loops with a certain configuration. We use the 
\textsl{depth-first search} (DFS) \cite{Sedgewick,Hartmann} algorithm to calculate all the connected components in the system. For example, 
in two-dimensional magnetic systems, the islands of magnetization are connected components of various sizes. The order-disorder transition 
in such systems is determined by the value of the magnetization order parameter. Unlike magnetic systems, the NS loops have no such order 
parameter. However, one can draw a close analogy with the idea of magnetization and the average number of loops of various sizes, as a measure 
of disorder in such loop systems. With a slight abuse of terminology, one may choose to call it a \textsl{disorder parameter}. 

The quantities that will be important for the present study are the \textsl{percolation probability} ($P^{c}$), and the \textsl{percolating cluster} 
($P_{\infty}$). In the previous section, we looked at the concept of a percolating path. If $N^{c}_{i}$ represents the number of times a percolating 
path of a given color $c$ occurs in a given direction $i$, out of $N_{M}$ trials, then

\begin{equation}
P^{c}_{i} = \frac{N^{c}_{i}}{N_{M}},
\end{equation}

is simply the percolation probability of color $c$ in the $i$-th direction. A percolating path is a connected component of a certain size which is 
determined by the number of lattice sites belonging to the percolating path. If $N^{c}_{\infty}$ represents the size of such a cluster of a specific 
color $c$, then we can define our second quantity of interest, the percolating cluster as

\begin{equation}
P^{c}_{\infty} = \frac{N^{c}_{\infty}}{L^{3}}.
\end{equation}

Where $L^{3}$ is the total number of lattice sites. It indicates the size of the largest cluster that spans the opposite boundaries of the system. 
This can be interpreted as the probability of a randomly chosen site being part of this cluster. Furthermore, it is an indicator of the random geometric 
networks created by the loops. 

In the upcoming section, we will observe that close to the critical region, these quantities exhibit scaling behavior. By using finite-size scaling 
analysis, we will determine the critical exponents associated with these quantities.

\section{Results}

\subsection{Finite-size scaling of the Percolation Probability}
\begin{figure}
\centering
\subfigure[]{\includegraphics[width=0.328\textwidth]{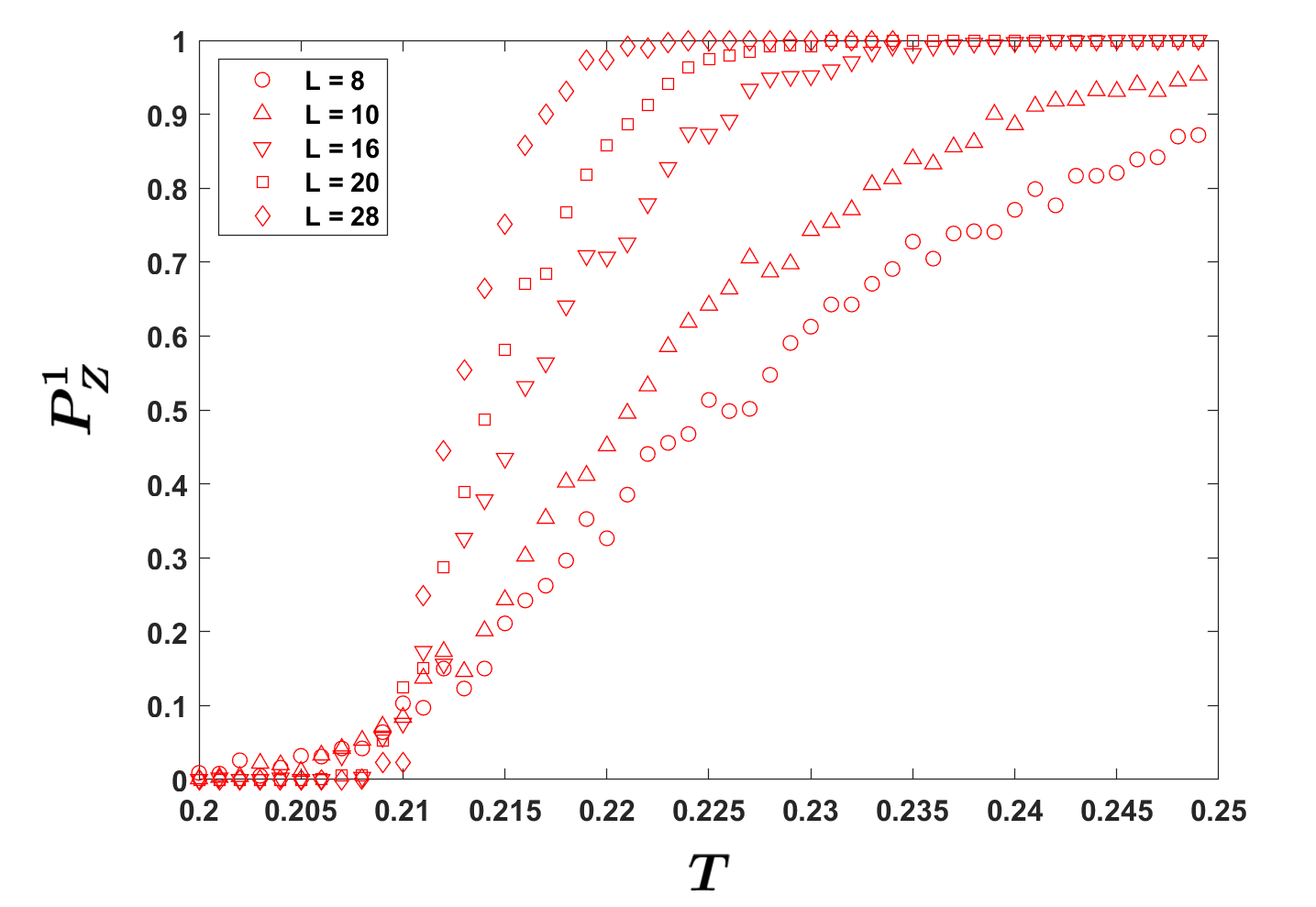}}
\subfigure[]{\includegraphics[width=0.325\textwidth]{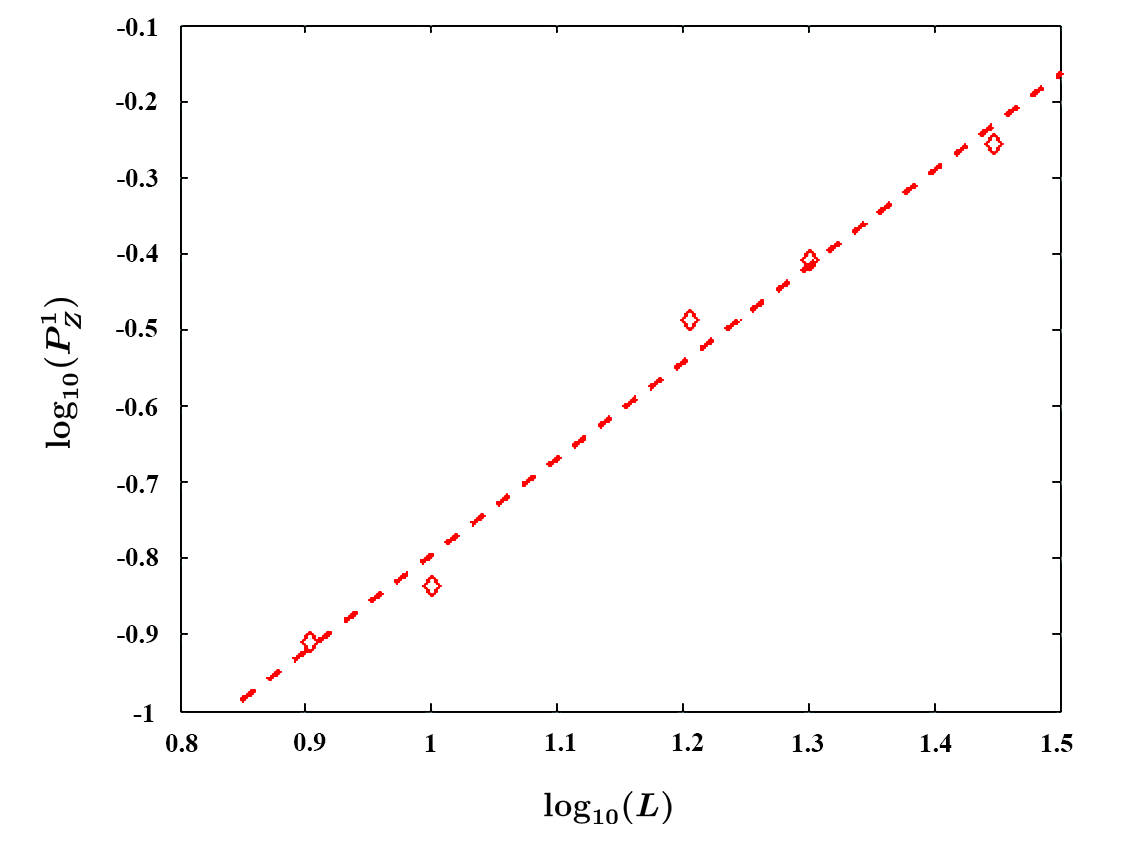}}
\subfigure[]{\includegraphics[width=0.327\textwidth]{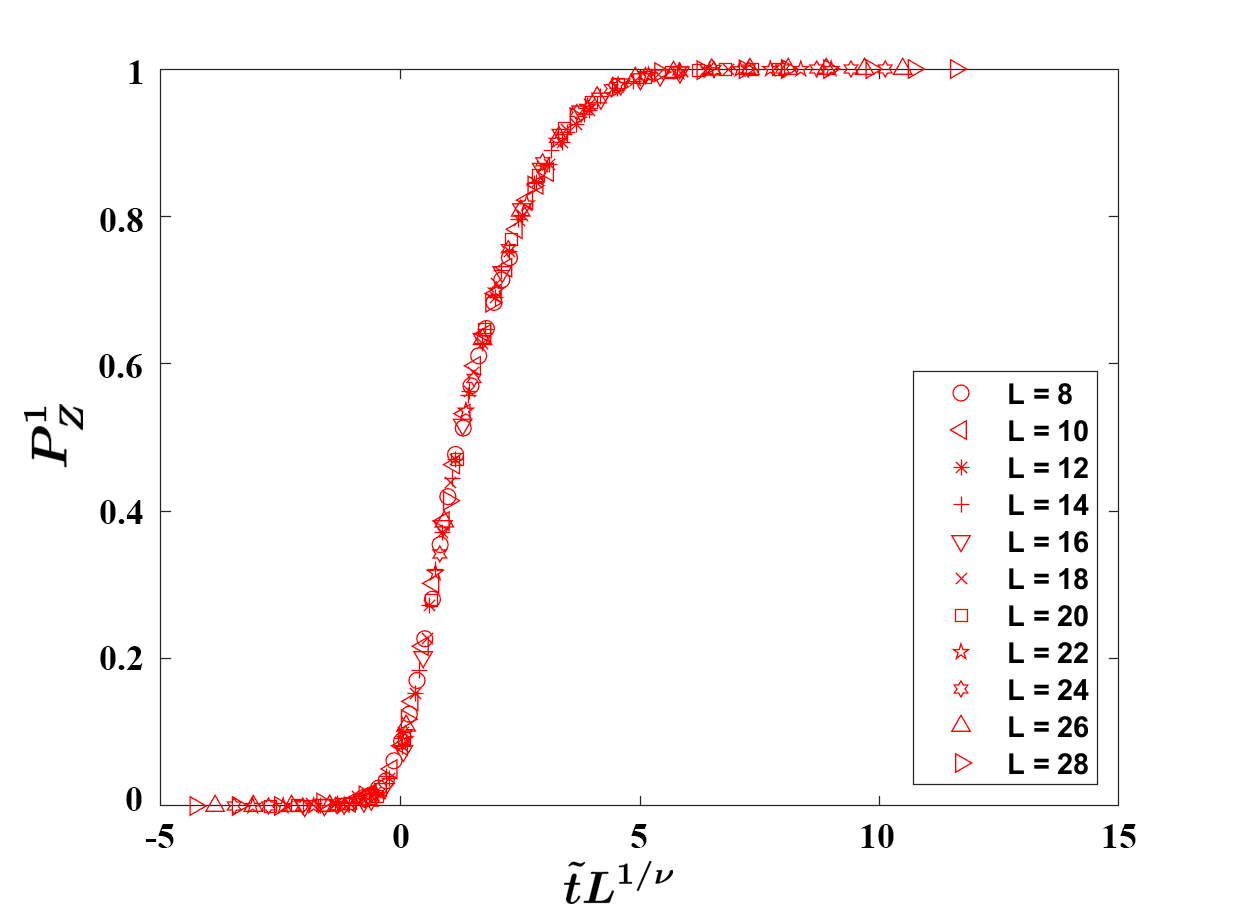}}
\caption{Percolation probability ($P^{c}_{i}$) in NS loops: 
(a) the raw data for $P^{1}_{Z}$ vs temperature ($T$); 
(b) $\log-\log$ plot for $P^{1}_{Z}$ vs $L^{\frac{1}{\nu}}$ for the correlation length exponent ($\nu$); 
(c) finite-size scaling results for $P^{1}_{Z}$ (for other colors, see TABLE.\ref{Correlationexponent}).} 
\label{PercolationProbabilityColor123}
\end{figure}

\begin{center}
\begin{table}
 \begin{tabular}{|| c c c ||}
 \hline
 & $T_{c}$ & $1/\nu$ \\ [0.5ex]
 \hline\hline
 $ P^{1}_{Y}: $ & 0.20(00971) & 1.36(005)  \\
 $ P^{1}_{Z}: $ & 0.20(00971) & 1.36(002)  \\
 $ P^{2}_{X}: $ & 0.20(00962) & 1.36(002)  \\
 $ P^{2}_{Z}: $ & 0.20(00973) & 1.36(002)  \\
 $ P^{3}_{X}: $ & 0.20(00978) & 1.36(002)  \\
 $ P^{3}_{Y}: $ & 0.20(00978) & 1.36(002)  \\
 \hline
\end{tabular}
\caption{Comparison of the critical temperatures ($T_{c}$), correlation length exponent ($\nu$) for different colors in different directions.}
\label{Correlationexponent}
\end{table}
\end{center}

As the system approaches criticality ($T \rightarrow T_{c}$), the loops grow in size and the system reaches the percolation threshold. Near this 
threshold point, the correlation length of an infinitely large system goes to infinity, but the correlation length is upper bound by the size of 
the finite-sized system ($L$). We find that with increasing system size, the $P_{i}^{c}$ vs $T$ curves become steeper near the percolation threshold 
as shown in Fig.\ref{PercolationProbabilityColor123}(a). A crude estimate of the critical temperature ($T \approx0.21$) may be obtained from the 
intersection point of these curves. However, to get a more quantitative and accurate result, we have performed finite-size scaling analysis of 
these loops of different colors using \textsl{autoscale.py} by Melchert \cite{Melchert} and our own numerical methods. Close to the critical point, 
$P^{c}_{i}$ has the following scaling form

\begin{equation}
P_{i}^{c}(L) = P_{i}^{c}(\tilde{t} L^{\frac{1}{\nu}}).
\end{equation}

The R.H.S. of the above equation is a homogeneous function of the scaled temperature $\tilde{t} L^{\frac{1}{\nu}}$, where $\tilde{t} = (T-T_{c})/T_{c}$. 
The exponent $\nu$ is associated with the correlation length (GMDa24)\cite{Ganguly}. Note that $P_{i}^{c}(L)$ becomes independent of the system size only 
at $\tilde{t} = 0$ (i.e. $T = T_{c}$). The correct choice of the exponent and the critical temperature $T_{c}$ will compel all the curves of different 
system sizes to collapse onto one another (Fig.\ref{PercolationProbabilityColor123}(c)). The critical temperature is shown in the 
TABLE.\ref{Correlationexponent} is close to $T_{c} = 0.21(0003)$. The correlation length exponent is approximately $\nu = 0.74(0024)$. The exponent 
$\nu$ calculated from the three-dimensional XY model gives $1/\nu = 1.48(0091)$ \cite{Vicari}.

\subsection{Finite-size scaling of the Percolating Cluster}

\begin{center}
\begin{figure}
\centering
\subfigure[]{\includegraphics[width=0.46\textwidth]{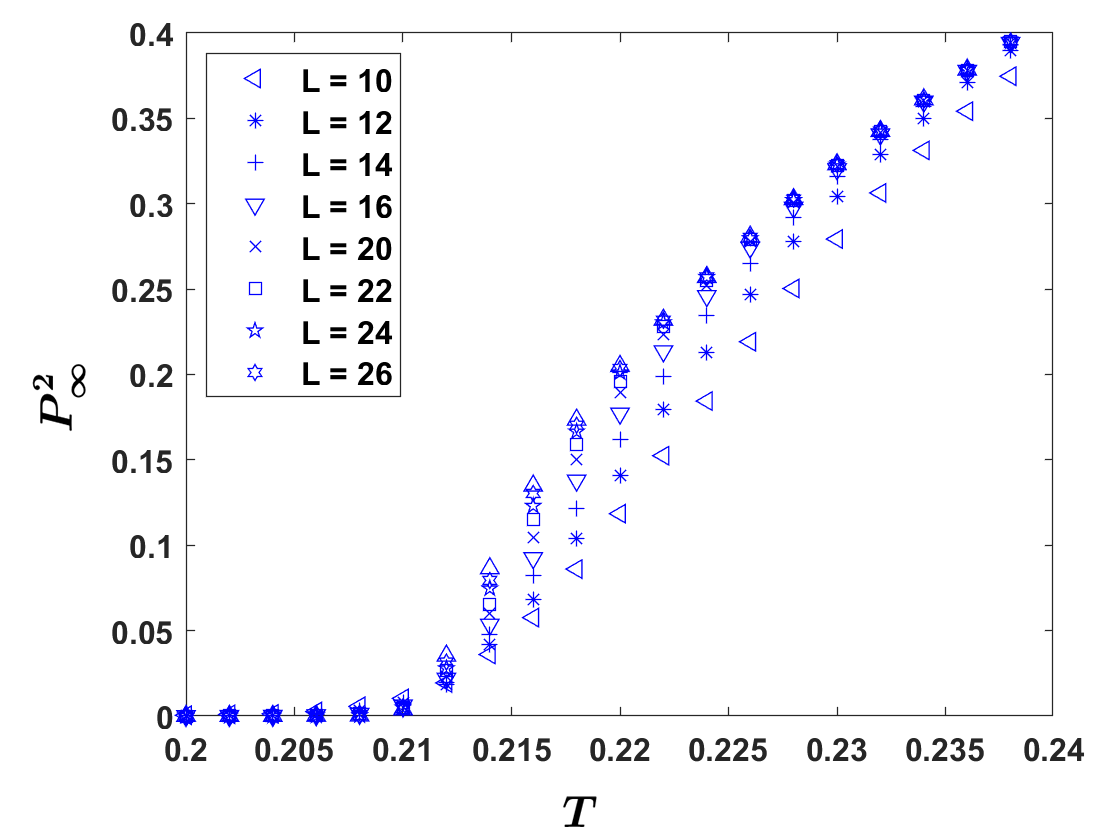}}
\subfigure[]{\includegraphics[width=0.46\textwidth]{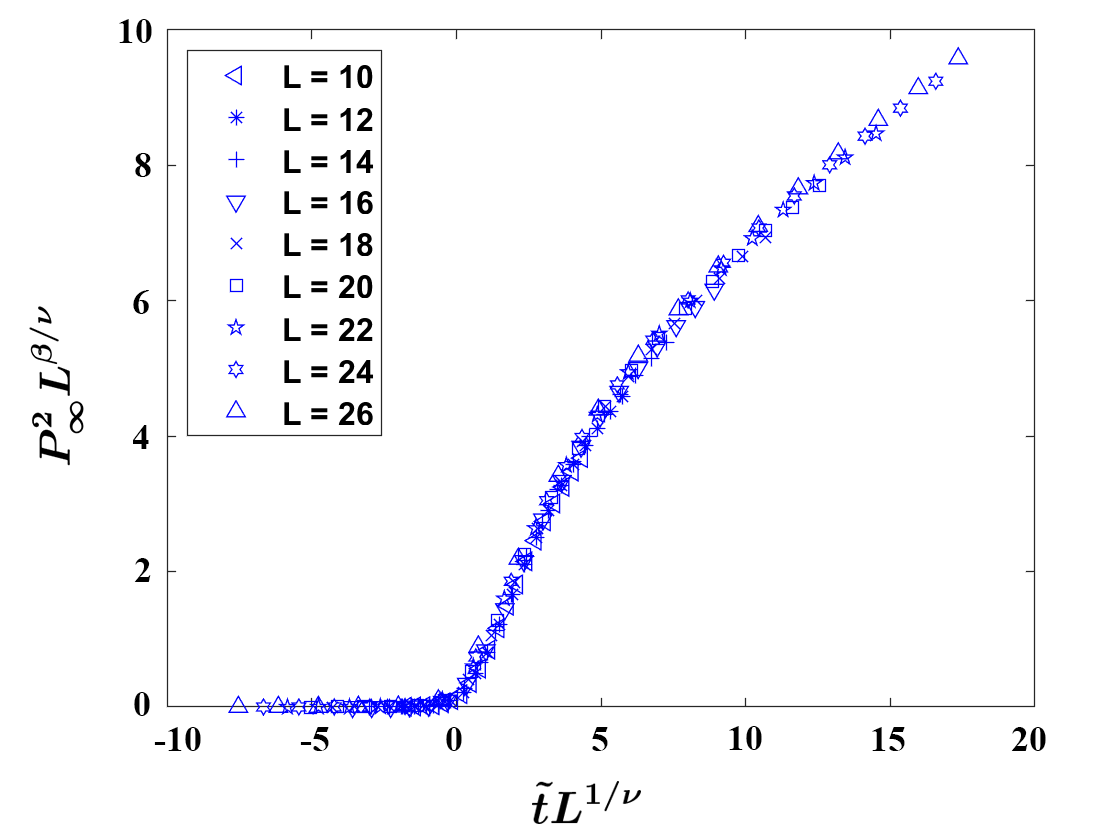}}
\caption{Percolating cluster $P^{c}_{\infty}$ in NS loops. Raw data and finite-size scaling results: (a) $P^{2}_{\infty}$ vs $T$ plot (color 2); 
(b) $P^{2}_{\infty}L^{\beta/\nu}$ vs $\tilde{t} L^{\frac{1}{\nu}}$ (for other colors, see TABLE.\ref{Percolating_Cluster_Exponent}).} 
\label{PInfinityColor123}
\end{figure}
\end{center}

Similar to the percolation probability, $P^{c}_{\infty}$ also has a scaling form given by

\begin{eqnarray}
P^{c}_{\infty}(L) &=& L^{-\frac{\beta}{\nu}} \bar{P}^{c}_{\infty}(\tilde{t}L^{\frac{1}{\nu}}),
\end{eqnarray}

where $\bar{P}^{c}_{\infty}$ is a homogeneous function of $\tilde{t} L^{\frac{1}{\nu}}$. For varying system sizes, $P^{c}_{\infty}$ has a behavior shown 
in Fig.\ref{PInfinityColor123}(a). Finite-size scaling analysis gives us the values of the exponents for various colors. The critical temperature 
obtained from the TABLE.\ref{Percolating_Cluster_Exponent} is close to $T_{c} = 0.21(0003)$, and the exponent $\beta = 0.61(007)$.

\begin{center}
\begin{table}
 \begin{tabular}{||c c c c c ||}
 \hline
 & $T_{c}$ & $1/\nu$ & $\beta/\nu$ & $\beta$ \\ [0.5ex]
 \hline\hline
 $ P^{1}_{\infty}: $ & 0.21(0003) & 1.31(0057) & 0.80(0056) & 0.61(002) \\
 $ P^{2}_{\infty}: $ & 0.21(0003) & 1.31(0057) & 0.82(0049) & 0.62(007) \\
 $ P^{3}_{\infty}: $ & 0.21(0001) & 1.31(0057) & 0.80(0056) & 0.61(002) \\
 \hline
\end{tabular}
\caption{Comparison of the critical temperatures ($T_{c}$), correlation exponent ($\nu$), and the exponent ($\beta$) for different colors.}
\label{Percolating_Cluster_Exponent}
\end{table}
\end{center}

\section{Conclusion}

Using Monte Carlo simulations, we have studied percolation in the Non-Symmetric loop model Eq.\eqref{NSLoopHamiltonian}.
The exponents are tabulated in Table \ref{Correlationexponent},\ref{Percolating_Cluster_Exponent}. The effects of
anisotropy (directional preference) are seen while the system transits from an ordered to a disordered state. The scaling
exponent for the percolation probability ($\nu = 0.74(0024)$) and percolation cluster ($\beta = 0.61(007)$) for NS loops
have been calculated. The percolation transition occurs at a temperature $T \approx 0.20(0098)$, which is close but not
exactly equal to the thermal critical temperature $T_{c} = 0.212$ (GMDa24)\cite{Ganguly}. A numerical study of the stochastic 
Gross-Pitaevskii equation by Kobayashi and Cugliandolo has shown that the percolation temperature of the vortices differed 
from the thermal critical temperature by only two per cent \cite{Kobayashi}. Similarly, a finite-size scaling calculation of 
the correlation length exponent of the three-dimensional XY model by Schultka et al.\cite{Schultka} and Vicari et al.\cite{Vicari} 
agrees with our model within a 6 per cent deviation. 

\section{Acknowledgement}

The authors acknowledge the financial support from the Council of Scientific \& Industrial Research (CSIR), the Indo-U.S. Science \& Technology 
Forum (IUSSTF), India. Also, the computational support from the Supercomputer Education \& Research Centre is acknowledged. S. Mukherjee and 
S.K. Ganguly thank Prof. N. Chandra in the Bioinformatics lab for partial computational support. C. Dasgupta thanks the Department of Science \& 
Technology (DST), Government of India. S.K. Ganguly would like to thank the entire ONGIL technical team, Mr. Amit Kumar Patra, Prof. Banibrata 
Mukhopadhyay, and Prof. Indrajit Mukherjee for their constant technical and moral support. Finally, we would like to express our gratitude to all 
the anonymous referees for their valuable suggestions.


\end{document}